# From Agents to Continuous Change via Aesthetics: Learning Mechanics with Visual Agent-based Computational Modeling

Pratim Sengupta
Amy Voss Farris
Mason Wright
Mind, Matter & Media Lab
Vanderbilt University - Peabody College

# From Agents to Continuous Change via Aesthetics: Learning Mechanics with Visual Agent-based Computational Modeling

## Abstract

Novice learners find motion as a continuous process of change challenging to understand. In this paper, we present a pedagogical approach based on agent-based, visual programming to address this issue. Integrating Logo programming with curricular science has been shown to be challenging in previous research on educational computing. We present a new Logo-based visual programming language - ViMAP - and, a sequence of learning activities involving programming and modeling, designed specifically to support seamless integration between programming and learning kinematics. We describe relevant affordances of the ViMAP environment that supports such seamless integration. We then present ViMAP-MoMo, a curricular unit designed in ViMAP for modeling kinematics, for a wide range of students (elementary - high school). The main contribution of this paper is that we describe in detail a sequence of learning activities in three phases, discuss the underlying rationale for each phase, and where relevant, report results in the form of observational data from two studies.

# Introduction

Novices find mechanics quite challenging to understand. Over the past three decades, researchers (Halloun & Hestenes, 1985; McColskey, 1983; McDermott, Rosenquist, & van Zee, 1987; Leinhardt, Zaslavsky, & Stein, 1990; Elby, 2000; etc.) have reported that novices (e.g., middle school, high school and undergraduate students) face conceptual difficulties in: a) understanding and explaining the formal (mathematical) relationships between distance, speed, time, and acceleration; and b) interpreting and explaining the physical concepts, relationships and phenomena represented by commonly used graphs of speed vs. time or distance vs. time. In particular, researchers have found that understanding continuous change in motion is challenging for novice learners. For example, when provided with a situation that involves objects moving with uniform acceleration (e.g., during a free fall or on an inclined plane), novices find it challenging to differentiate between instantaneous speed and average speed, (Halloun & Hestenes, 1985) and tend to describe or explain any speed change in terms of differences or relative size of the changes, rather than describing speeding up or slowing down as a continuous process (Dykstra & Sweet, 2009).

Prior research on programming based pedagogical approaches for teaching and learning physics in K12 classrooms suggests that agent-based programming —in particular, programming with Logo based languages, where learners control the behavior of a protean computational agent (e.g., the Logo turtle) by using simple programming commands — can serve as an effective pedagogical approach to help students in middle grades learn about kinematics through modeling. (diSessa, Hammer, Sherin & Kolpakowski, 1991; Sherin, diSessa, and Hammer, 1993). However, a challenge of using Logo-based programming languages for teaching physics is that it involves extensive programming instruction and scaffolding, in addition to teaching physics. This in turn can led to a very high overhead for teachers, as Sherin, diSessa, and Hammer (1993) pointed out.

Our aim in this paper is to present a solution to this challenge of integrating programming and computational modeling seamlessly with classroom physics instruction in the particular domain of kinematics. To this end, we present ViMAP, a new Logo-based visual programming language and modeling platform, and ViMAP-MoMo (Modeling Motion with ViMAP), a curricular unit in kinematics based on agent-based programming and modeling activities designed in ViMAP. We discuss how we designed the programming language and learning activities so that students can develop conceptual understandings of motion as a continuous process of change by constructing mathematical relationships between distance, speed, and acceleration using turtle graphics.

Turtle graphics, also called turtle geometry, has a long history in Logo-based learning (e.g., Abelson & diSessa, 1981). By commanding the movement of the Logo turtle on the computer screen using body-syntonic programing primitives, Papert (1980) and Abelson & diSessa (1981) showed how students can explore the properties of space by following the turtle's actions. We demonstrate how variations of this genre of Logo-based activities can be adapted to teach key mathematical relationships central to understanding motion as a process of continuous change, by supporting the development of relevant scientific representational practices (e.g., modeling and graphing) at the elementary and high school levels. Our proposed pedagogical approach does not necessitate a high programming overhead; rather, has an explicit emphasis on aesthetics. As students gradually develop and design models as selective abstractions of phenomena, they intuitively attend to aesthetic relationships, i.e. space, structure, and symmetries of the turtle graphics that they generate using visual programming. As Weschler (1978) suggested, we see the aesthetic considerations that are taken up in the processes of modeling as potentially deterministic to the "form, development, and efficacy of models" (p.3), much akin the aesthetic judgments that guided the development of scientific theories (Weschler, 1978; Miller, 1978).

Our main focus in this paper is on the design of ViMAP and ViMAP-MoMo, rather than providing extensive empirical support. That is, our goals are to highlight the relevant elements and characteristics of ViMAP and to describe a sequence of learning activities with an explicit emphasis on aesthetics that we believe can help integrate agent-based programming with classroom physics in a seamless manner. These elements and characteristics include the following: a) design of both domain-specific and domain-general visual programming primitives, b) scaffolds for debugging and making the user's code "live" (Tanimoto, 1990), and c) integrating graphing within the programming environment. The sequence of learning activities we describe here consists of three distinct but related phases. We discuss the rationale behind the design of these elements and characteristics of ViMAP, as well as the learning activities. In addition, where relevant, we also provide some empirical support for some of these activities, based on qualitative observations from studies we have conducted with elementary and high school students.

## Background

**Learning Physics with Logo**

### Logo Beyond Programming

One of the earliest research groups to develop Logo-based programming and modeling environments for K12 children was the Berkeley BOXER group (diSessa, Hammer, Sherin, and Kolpakowski, 1991; diSessa, 1985; diSessa & Abelson, 1986). As diSessa, Abelson & Polger (1991) pointed out, Boxer was designed in response to two drawbacks of Logo in terms of classroom use: a) although simple procedures are easy to generate in Logo, mastering the language is quite challenging; and b) while many teachers, after extensive experience with Logo, could create simple computer-based microworlds consisting of a few Logo procedures, Logo's lack of structuring principles, beyond individual procedures, made it difficult for teachers to organize these into flexible constructs, such as an interactive notebook that students can use and modify. In other words, adapting the general-purpose nature of Logo for domain-specific applications in classrooms was challenging.

To address these issues, Boxer was based on two principles of learnability: concreteness and spatiality. Concreteness specifies that all mechanisms in the system should be visible and directly manipulable on the display screen. The second underlying principle is a uniform spatial metaphor for structure. The root form of Boxer is an object called a box, which may contain text, graphics, programs, or other boxes. The spatiality of boxes allows people to use ordinary spatial intuitions of inside, outside, and next to in order to understand a broad range of computational structures. An important characteristic of Boxer is that the student-generated program is part of the same environment as the enacted output. That is, students can make changes to the underlying rules governing the motion of the turtles and see the resultant output in the same visible software environment.

**Challenges in Integrating Boxer Programming with K12 Physics Curricula**

As Guzdial (1994) pointed out, the challenge in using programming to learn other domains is that programming as an activity often requires more skills and knowledge which are disconnected from the domain-specific learning goals. Boxer was designed to address some of these challenges for Logo-based, domain-specific learning, as we have discussed in the previous section. However, our review of the literature reveals that integration of Boxer-programing with classroom physics was still found to be challenging. This is due to the fact that in order to program computer simulations of any physical phenomenon, both teachers and students must have some operational fluency with the programming language being used for instruction, in addition to the relevant domain knowledge. Therefore, in classroom-wide research studies of physics curricula that involve programming, the curricular units devote a significant amount of time on programming instruction prior to science instruction. In the studies reported by Sherin et al. (1993) and diSessa, Hammer, Sherin, and Kolpakowski (1991), middle and high school students underwent 15 weeks of instruction, out of which the first five weeks of classroom instruction were devoted solely to learning programming taught by a programming expert. As Sherin, et al. (1993) pointed out, the additional overhead of teaching programming "may simply be prohibitive" (p. 116) when considered along with the time constraints faced by science (in their case, physics) teachers,

## The ViMAP Learning Environment

ViMAP is a new multi-agent-based visual programming language (Sengupta, 2011; Sengupta & Wright, 2010) based on the NetLogo modeling platform and environment (Wilensky, 1999). NetLogo is a multi-agent based modeling environment in which the user can create, interact with, or manipulate thousands of agents, whose behaviors are controlled by simple rules that can also be specified and controlled by the users. NetLogo is in widespread use in both educational and research contexts, and a variety of K16 STEM curricula have been developed in the NetLogo environment. In this paper, we present a version of ViMAP, called ViMAP-MoMo, which includes programming commands and domain-specific microworlds that are specifically designed for modeling kinematics.

ViMAP is a multi-layered, flexible programming language and modeling environment that can be modified and extended easily for domain-specific use. That is, at one level, using ViMAP, novice programmers can construct a new simulation or modify an existing simulation using visual and tactile programming, usually known as visual programming (or visual modeling). At another level, ViMAP takes advantage of the glassbox nature of NetLogo (Tisue & Wilensky, 2004) and allows any moderately experienced Logo programmer to extend the ViMAP language by adding new programming commands in the underlying NetLogo language. The version of ViMAP we present in this paper can be run either locally on a desktop or as an online applet.

The design of ViMAP builds on the previous research on Logo-based modeling environments such as Boxer, and takes significant further steps in integrating agent-based programming and modeling with learning kinematics. Like Boxer, ViMAP also utilizes the principles of concreteness and spatiality. But it extends the principle of concreteness by enabling students to a) see the program at the same time as the running environment. The program even updates the relevant parameters as the system runs; and b) generate graphs within the same environment and toggle back and forth between the graph and the ViMAP code. Students are thus introduced to multiple computational and mathematical representations of the simulated phenomenon within the same learning environment. The principle of spatiality is extended by using visual programming (as opposed to text-based programming) as the mode of construction of algorithms, as well as generation of graphs, which we describe in the section on the sequence of activities. When students construct ViMAP algorithms (e.g., Figure 2b), they can use their intuitions of spatiality to arrange the programming commands by dragging and dropping them in the appropriate spatial order. The graphing functionality, which we describe in section detailing the sequence of the learning activities, allows the student to generate measured bars, corresponding to particular elements of the wabbit's enacted motion, and arrange them in a temporal order using mouse-clicks and a drag and drop interface (e.g, Figure 8).

Another important affordance of ViMAP is the combination of domain-specific and domain-general programming commands. Besides using domain-general commands for control flow, the ViMAP language also contains domain-specific programming commands such as speed-up and slow-down. This is discussed further in the following sections, when we present the activities in more detail. We believe that this plays an important role in reducing the overhead involved in learning Logo programming in a physics classroom.

The ViMAP world is divided into two parts: construction-world (C-World), where learners construct their own programs; and enactment world (E-World), where a protean computational agent (or a set of agents)—the classic Logo turtle (or a set of turtles)—carries out learners' commands through movement in representational space. The programming commands in the C-World are also represented visually as a class of NetLogo turtles, which in turn can be dragged and moved by mouse-clicks. Figure 1 shows all 23 programming commands available to users. To minimize confusion between enacting turtles and command block turtles, we have termed the enacting agents "wabbits" and the other class of agents "command blocks" or "code blocks," terms that we use in the rest of this paper. The learner can construct her program by selecting commands from a list, choosing initialization values with sliders,(when the command has a parameter)and spatially ordering commands using a visual, drag-and-drop, snap-to-grid[1] interface. ViMAP programming commands include primitives that control wabbit motion, interactions among wabbits, and control flow. These commands can be used to produce simple animations through pen stroke and clear-all commands, to model physical interactions between wabbits using commands that control wabbit speed, direction, and acceleration, and to place flags that are used in measuring distances within the model.

In the unit we describe in here, learners begin with a single-agent version of ViMAP and then progress to a multi-agent version. In the single agent version, learners can control the behavior of a single wabbit in the E-world, and in the multi-agent version, learners control the behaviors of multiple wabbits at the same time. In the multi-agent version, separate, tabbed pages of command blocks for different wabbits (or classes of wabbits) allow the user to program each wabbit (or class of wabbits) with distinct code sequences. When the user clicks "Run Once" to execute the program, ViMAP randomly shuffles an array of all wabbits in the E-world, then asks each in turn to execute its complete sequence of commands. The user can choose for each wabbit to run its agentset's code once (i.e., for one iteration only) or for the code to run in an endless loop.

Learning with ViMAP involves rapid cycles of construction, execution, and refinement of programs that control wabbit behaviors based on feedback from the execution. To support

---

[1] Each line of code in the C-World can accommodate a maximum of six command blocks

learning activities that involve rapid prototyping, ViMAP offers learners a range of "liveness" factors (Tanimoto, 1990) for algorithm visualization. For example, as commands are being executed, the currently active agentset's tabbed page of code blocks is brought to the foreground, and the currently running code block is highlighted in red. The user can also set a time delay between individual command blocks. As a further aid debugging and code visualization, learners can choose to edit their code while the simulation is running, or they can choose to stop the simulation, examine the output,compare it to the phenomenon they are modeling, make appropriate changes to the code, and then re-rerun the simulation. Learners can also choose to overlay the graphical results (in the E-World) from multiple runs, and the results of each run can also be saved in a different color.

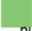

Figure 1. A List of ViMAP Commands with descriptions of functions. Within the Figure, "n" is used to represent a parameter selected by the user using a slider.

**Walkthrough: A Typical ViMAP-MoMo Activity**

To give the reader a vivid sense of how ViMAP works, we will now describe a typical activity in which a hypothetical user first generates a rectangular spiral, and then produces a graph representing the lengths of the consecutive sides of the spiral. Figure 2b shows the initial ViMAP code that generates the spiral.

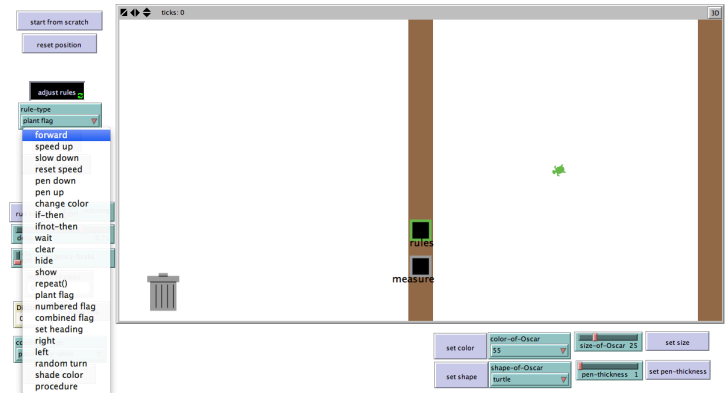

Figure 2a. User selects ViMAP Command from the Command List

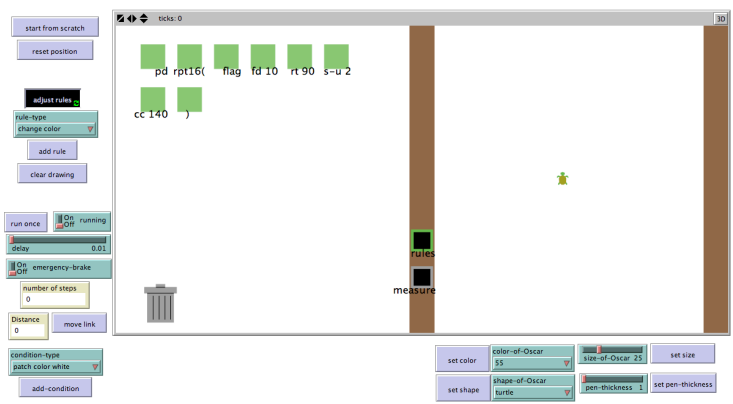

Figure 2b. ViMAP Code for Generating a Flagged Rectangular Spiral Growing Outward

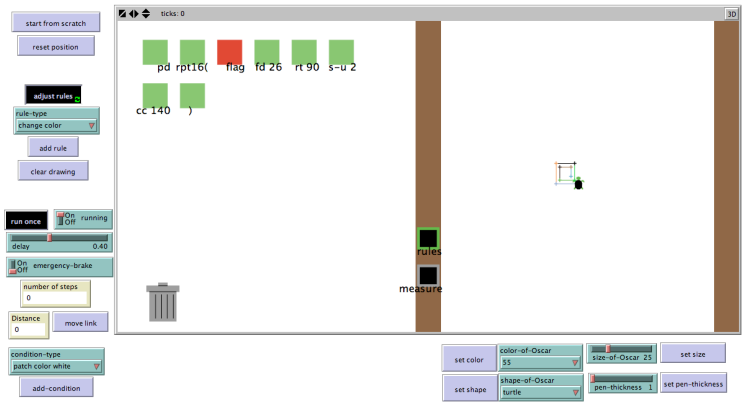

Figure 2c. Screenshot of Execution of the ViMAP Program in Figure 2a.

As shown in Figure 2a, the user selected the ViMAP commands by clicking on the drop-down menu titled "rule type", which then displays a list of available ViMAP commands. The user then composes the ViMAP algorithm by selecting the relevant commands, pressing "add rule", and then ordering the commands spatially in the C-World to produce the code as shown in Figure 2b. In this case, the user's code first asks the wabbit to put its pen-down, which in Logo, means that the wabbit will create marks on the screen continually as it moves, unless it is asked to stop doing so by using the pen-up command. Next, the user's code asks the wabbit to generate a rectangular spiral by using commands for planting a flag (indicated on the screen by a crosshair), moving forward by a certain number of steps (forward 10), turning right by 90 degrees (right 90), and then moving forward again by a magnitude that exceeds the previous step-size by 2 units (speed-up 2). The user's code asks the wabbit to repeat these steps, in order, for a total of sixteen times (repeat 16). Figure 2c shows a screenshot taken during the eighth iteration of the program, and the command currently being executed in the E-world is highlighted in red.

**Sequence of Activities in ViMAP-MoMo**

In this section, we present a detailed discussion of a sequence of learning activities that we designed for ViMAP-MoMo.    The sequence of activities consists of three phases and each phase consists of multiple activities. We also present the rationale underlying the design of each phase. Furthermore, pertaining to some of these activities, we present some data in the form of qualitative observations from two studies that we have conducted in the past couple of years. Study 1 (Sengupta & Farris, 2012) was conducted in a large metropolitan city on the campus of a large private university in the mid-southern USA. Fourteen children in 3rd and 4th grades at local schools were recruited by email and web solicitation for a six-session weekend course, and classes met once a week for two and a half hours, on six consecutive Saturday mornings. Study 2 (Sengupta & Hubbell, submitted) was conducted in a 100% African American, public charter school in a large mid-south metropolitan area with seven students in the ninth grade. At the time the study was being conducted, the school was in a probationary period due to low academic performance. The study spanned eight one-hour sessions, which took place in the school's computer lab and were taught by a member of our research lab who had some prior familiarity with ViMAP, but was not involved with the development of ViMAP.  Seven students participated in the learning activities. In both studies, none of the students in this course had any prior programming experience.

**Phase 1: The Physics of Aesthetics**

    **Design Rationale**

Our design of the activities was guided by the following rationale: 1) Aesthetics and visualization are robust points of entry that can be leveraged for learning mechanics; and 2) Students will develop familiarity with programming using ViMAP primitives. This involves using both the domain general primitives for control flow, as well as the domain specific commands for movement. The graphical representations being generated in the E-World are non-canonical—i.e., the process of generating geometric shapes is not traditionally used to teach students about kinematics. However, given that the shapes are familiar to students, the outcomes of the problems are familiar to them; the new learning here happens in translating the act of drawing a simple shape in terms of the ViMAP commands for turtle movement (which are explicitly physics based), and control flow. The programming commands that students use in this phase will also form the core of their models in later phases, where they use these shapes to mathematically represent motion in the real world. This tight coupling between learning programming and learning physics is one of the central design principles of ViMAP (Sengupta, 2011). Phase 1 can therefore be understood as a set of activities that fosters necessary competencies (e.g., agent-based thinking, proficiency with programming including syntax, commands and control flow; and debugging) to learn physics using computational modeling and thinking in ViMAP.

Our emphasis on aesthetics goes beyond the obvious crowd pleaser: creating visually pleasing computational designs. Our emphasis also includes the mathematical form of the generated graphic, which as the historian of science Arthur Miller (1978) pointed out, played important roles even in the development of theories on physics. This can be understood as follows. The acts of modeling motion and drawing are indistinguishable from one another in ViMAP because a wabbit with its pen down leaves behind an inscription (marking) of where it has been. Shapes emerge from these markings over time, as the model runs over multiple iterations. These shapes render a concrete form to the student's code. This in turn makes explicit the mathematics underlying the ViMAP code, which in turn represents the target physical mechanism. For example, the simple act of generating a straight line over three successive iterations, where each segment of the line is of a different color, can make explicit the amount of distance travelled by the wabbit during each interval of time. Thus, the aesthetic attention that students give to their work is not distal from their attention to modeling motion. On one hand, changes of color or spatial properties of the generated graphic (e.g., symmetry; space between lines; angles; etc.) may be made as a function of whimsical (or playful) artistic expression, and this may also be a personally meaningful activity for the student. However, on the other hand, it is also important that the aesthetic choices that children make in their modeling and programming have mathematical and communicative values about the underlying mathematical and physical assumptions. Previous studies on modeling show that student-created models can become "circulating references" (Latour, 1999) in the classroom, leading to collective disciplinary engagement (Lehrer, 2009); therefore, students' aesthetic choices, when coupled with the domain-specific design goals, can make explicit what is worth noticing in their models in terms of the underlying kinematics and mathematics.

**Activities**

*Activity 1.1: Generating "Constant Speed" Shapes*

In this first activity, students are briefly introduced to the environment and asked to program a single wabbit to draw a square, a triangle, a circle, and a stick figure. In doing so, students familiarize themselves with the use of available programming commands ("forward", "right turn", "left turn", "pen down", "set heading", "pen up" and "repeat"), and debugging.

Figure 1 depicts a composite image of ViMAP programs and resultant shapes, all generated within the constant speed paradigm.

Results from both the studies show that these constant speed shapes play an important role in the development of students' understanding of two key aspects of computational thinking: learning to think like a turtle (or agent-based thinking), and debugging. One of the common challenges faced by students for this activity involves identifying the correct "heading" of the Logo turtle , and use of commands for control flow, such as "repeat". Debugging, in turn, involves two components: first, identifying which elements of their programs correspond to which aspects of the output, and second, iterative refinement of their program so that it generates the desired or target output. Both the studies also reveal that teacher-led scaffolding in this stage is required primarily to prompt learners for agent-based thinking, for introducing students to commands for control flow and the relevant software-embedded scaffolds for debugging reported earlier in this paper (Sengupta & Farris, 2012; Sengupta & Hubbell, submitted). We found that an example of an effective prompt for agent-based thinking was asking students to act like a turtle, i.e., physically carry out the commands themselves by moving their own bodies according to the ViMAP commands they composed. Furthermore, we found that students would often benefit from using the "change color" command after every step of the turtle's motion, as it made the actions of the wabbit during each step distinct from the others. This, in conjunction with the code-stepthrough functionality, can enable students to develop a deeper understanding of the relationship between their ViMAP code in the C-World, and the enacted output in the E-World.

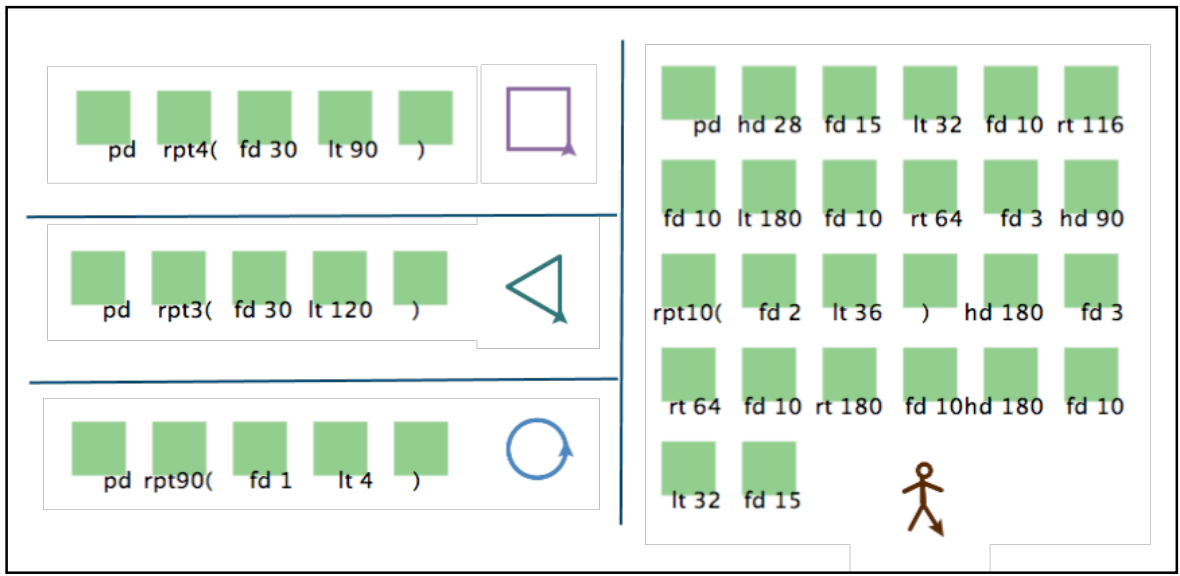

Figure 3. Possible solutions for "constant speed" shapes

*Activity 1.2: Generating "Constant Acceleration" Shapes*

In the second activity, students are asked to use the "speed up" and "slow down" commands, in addition to the other commands they previously used, to create two types of spirals: spirals that "grow" and spirals that "shrink". This activity is designed to be a conceptual stepping stone to Phase 2, where students will use these acceleration-based commands in order to model the motion of moving objects existing outside of ViMAP. Figure 4a shows images of spirals (two different E-Worlds) generated by a 9th grade student in one of our studies (Sengupta & Hubbell, submitted), and Figure 4b is a screenshot of a sample ViMAP code (in the C-World) that generates a rectangular spiral growing outward (in the E-World).

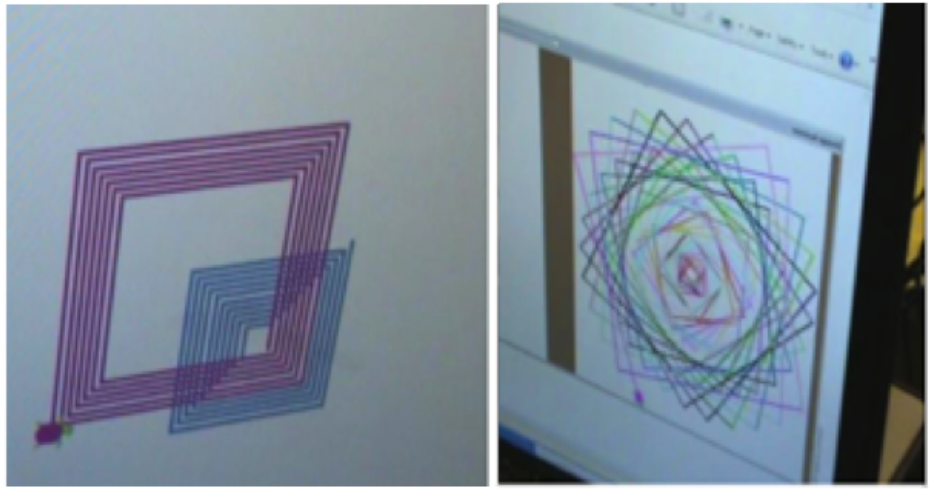

Figure 4a. Spiral patterns generated by a 9th grade student. The image on the left shows two rectangular spirals, and the image on the right shows a rotating rectangular spiral.

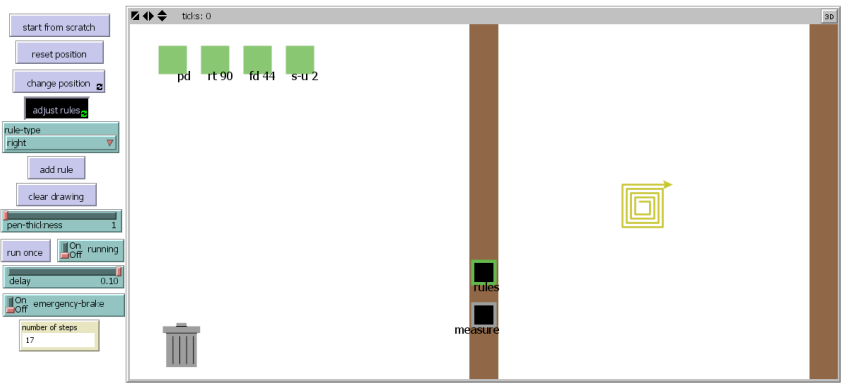

Figure 4b. Sample code for a rectangular spiral growing outward.

The code as shown in the C-World of Figure 4b is a sequence of four simple commands: pen-down (pd), right turn (rt), forward (fd), and speed-up (s-u). Note that the numerals displayed next to rt represent the angle of rotation, while the numeral displayed next to fd indicates the step-size of the turtle during the current (17th) iteration. The current iteration number is displayed in the "Number of Steps" monitor. This coupling between programming commands and parameters was designed specifically as a scaffold to focus learners' attention to the changing values of the forward command, which changes exactly based on the magnitude of acceleration displayed in the speed-up command block. This in turn makes explicit the instantaneous relationship between speed and acceleration. The C-World, therefore, explicitly displays the wabbit's current state using three elements:the current step-size is displayed next to the forward command, the amount of constant acceleration is displayed next to the speed-up command, and the number of steps taken by the wabbit is shown in a monitor.

Our studies show that this activity can enable learners to formally articulate their intuitive understanding of a continuous process of change in terms of repeated, discrete increments or decrements of step sizes. For example, in Study 1 (Sengupta & Farris, 2012), we found that a student (Nathan) already understood the mathematical mechanism for generating the rectangular spiral,and created a series of shapes with decreasing side lengths (not spirals) by iteratively reducing the forward step-size and re-running his code manually . He had not yet discovered the functionality of the "slow-down" command, or mastered the use of the "repeat" functionality, and was therefore experiencing challenges in implementing his qualitative understanding in a formal manner using ViMAP commands. However, it is in attempting to represent the mechanism formally, that he was able to debug his earlier program with assistance from the instructor. The assistance that the instructor provided was in the form of pointing out the "slow down" command and showing Nathan how to use two liveness scaffolds – the delay functionality, and the code step-through highlighter. It was by using these software scaffolds that Nathan was also able to identify the mathematical relationship between the commands "slow-down" and "forward" and successfully generate the spiral.

*Activity 1.3: Projections of Three Dimensional Figures*

The observable qualities of objects encompass data from three visuospatial dimensions, which we usually represent on flatlands (Tufte, 1990), or 2D surfaces. We have found that escaping this flatland through designing 2D projections of 3D objects can also serve as a highly engaging activity for young learners. Through these activities, learners can explore ideas of speed and acceleration in a non-canonical yet meaningful manner. Using ViMAP commands, students can create perspective drawings similar to the one shown in Figure 5 by speeding up and slowing down the ViMAP wabbit. In this example, the figure may be interpreted either as a series of 2D regular hexagons or a stack of 3D cubes. A relatively short list of commands is needed to create this fairly complex representation, as shown on the left side of Figure 5 (which was generated by one of the authors).

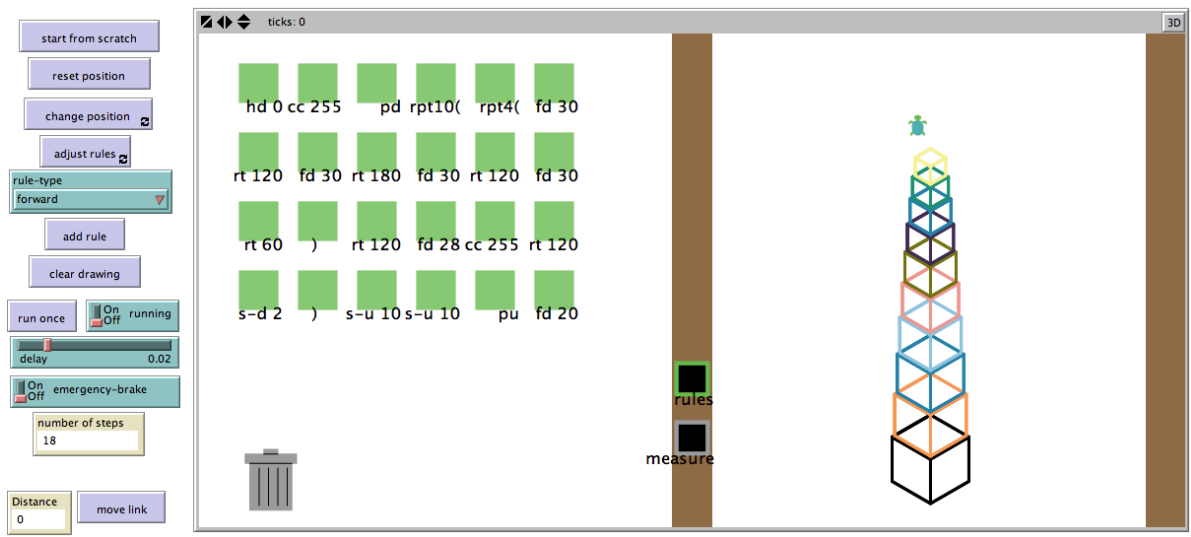

Figure 5. Overlapping hexagons or a stack of square prisms

*Activity 1.4: Painting with Turtles*

Previous research shows that creative expression can be a productive and personally meaningful route towards computational literacy (Eisenberg & Beuchley, 2008). Instead of using a physical paintbrush, in this activity students use ViMAP commands to generate turtle drawings. In this coupling of art and computation, students harness the computational power of the medium while being engaged in a personally meaningful activity. In one of our studies, we found that after students mastered some expertise in using commands for constant speed and constant acceleration to draw the canonical geometric shapes discussed earlier, they spontaneously began to draw new shapes and paintings that were personally meaningful to them (Sengupta, 2011). For example, in the tree picture in Figure 6 (which was generated by one of the authors), the repeating lines of the trunk and the overlapping whorls of the foliage are computationally created by a single wabbit, which moves forward, changes heading, and then either speeds up or slows down after every iteration. Results from both the studies show that as students create paintings in ViMAP, they become increasingly proficient in thinking like a turtle, using ViMAP commands, as well as in debugging (Sengupta & Hubbell, submitted; Sengupta & Farris, 2012).

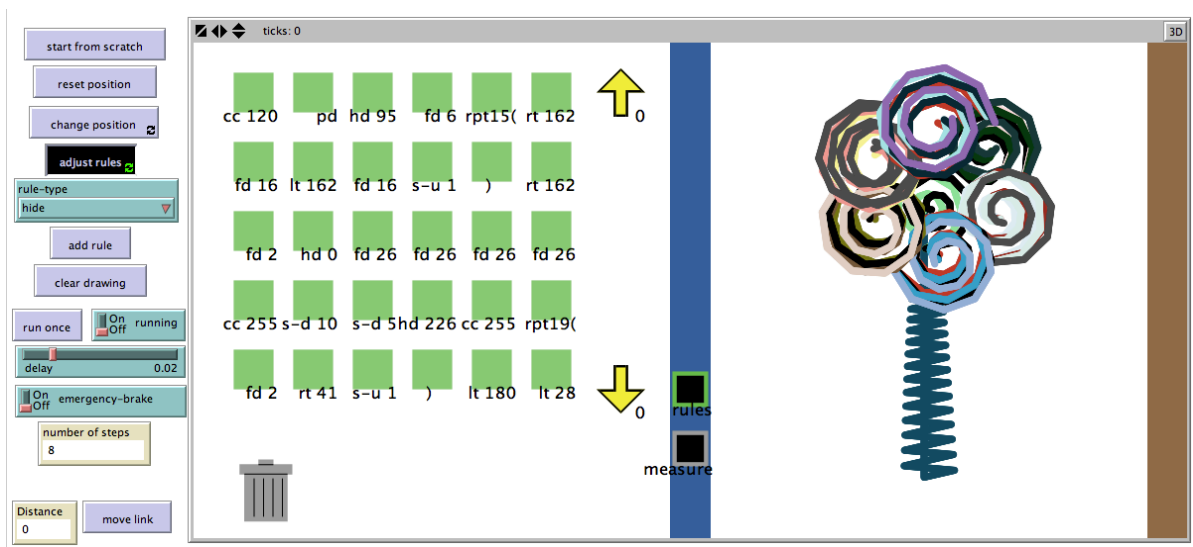

Figure 6. "Turtle art" representation of a tree

**Phase 2: Modeling Real-World Motion**

   **Design Rationale**

   In Phase 2, students are asked to use the same primitives they used in Phase 1 to model "real-world' motion involving a physical setup of two balls rolling down two ramps of different elevations and one ball being pushed on a horizontal surface (the floor). Our goal here is to introduce students to key aspects of scientific modeling (Hestenes, 1993). These aspects include the following: a) Observing initial setup conditions: Students learn to observe carefully the relevant conditions in the real-world phenomena that in turn will guide the development of the computational model, e.g., noticing that both the balls on the ramps start from rest, while the ball on the floor is pushed initially; b) Model construction through iterative refinement, analysis and validation: Scientists develop models through an iterative process that involves gradual refinement of the underlying assumptions, as well as the representational structures and systems being used, through a process of repeated comparisons between the model and the phenomena being modeled (Nercessian, 1992). Therefore, in addition to learning to model kinematic phenomena, students also learn to analyze the structure or implications of a given model, and to evaluate the capability of a model to account for given data or describe/explain given concrete properties and events.

   **Activities**

   *Activity 2: Modeling real-world motion*

Students are asked to model the movement of three balls moving with different values for acceleration. Two of the balls start from rest and roll down two ramps at different angles from the floor, and the third ball is initially pushed on the floor by any given amount of force. The real-world scenario can be presented in the form of a pre-recorded video, or learners can enact the motion themselves when provided with planks of wood and plastic balls (or non-motorized Lego cars).

This activity involves a modified and more case-specific version of ViMAP-MoMo. In the E-World of this environment, there are three wabbits, each represented as different colored balls. Each of these balls is on a ramp of different slope. The initial setup for ViMAP-MoMo is designed to model the scenario of the racing balls from the first activity (Figure 7). The blue ball rests on a flat plane, and all "forward" commands impel the ball along this plane. The yellow and red balls each sit atop an incline, and all "forward" commands impel the balls down their respective incline. The students are asked to create three separate sets of programs, one for each ball. They can toggle between the sets of code for each ball by clicking on the corresponding tabs (square icons) titled "red ball", "yellow ball" and "blue ball", located in the center of ViMAP-MoMo world (Figure 7). The goal of the students is to program each ball so that the simultaneous movements of the three balls simulates the motion of the real balls (or cars) in the physical setup (or the video).

An important element of ViMAP-MoMo introduced in Phases 2 and 3 is the plant flag command. A long-supported Logo-based representation of motion in introductory physics is the dot-trace representation (diSessa, Hammer, Sherin & Kolpakowski, 1991; Sherin, 2000). ViMAP-MoMo uses a flag (a cross-mark) in lieu of a dot as a representation of instantaneous position. Imagine a simple program in which the wabbit moves forward by certain distance during each iteration, and plants a flag at the completion of each iteration. The distance between two successive flags provides the learner with a measure of displacement during each iteration (or time interval). If one considers each iteration to be a unit interval of time, then this displacement also becomes a measure of the instantaneous speed of the wabbit. The learner can determine the total time elapsed by counting the number of flags.

The distances between the flags can reveal overall trends of the wabbits' motion in time. As shown in Figure 7, while both the red and yellow balls start from rest, they experience different amounts of acceleration—the red ball travels a steeper slope and therefore accelerates faster. This becomes evident in the relative distances between two successive flags for each ball: the flags of the red ball are farther apart than those of the yellow ball.

In both of our studies, we found that students need scaffolding in this phase. While conducting this activity, students should be frequently asked to compare the actions of the blue, red, and yellow balls in the real-world and about how these differences would show up in their programmed simulations. In trying to model a real world scenario through such frequent comparisons between the modeled and the real worlds, our studies show that students in elementary grades (Sengupta & Farris, 2012) and in high school (Sengupta & Hubbell, submitted) learn to recognize important constraints in the scenario to be modeled. For example, initial conditions play an important role in kinematics and mechanics, and in this case, in order to model the scenario shown in the video or the physical setup, students must realize that the yellow and red balls should start from rest. This process of model refinement and verification helps students realize that a ViMAP-MoMo model where the red ball reaches the end first is not necessarily a complete model, and that even when the red ball may appear to be behind the yellow ball for the first few moments, it will still win in the end. The blue ball, on the other hand, is slowing down, as evident in the decreasing distance between successive flags in Figure 7. In Study 1 (Sengupta & Farris, 2012), for example, in a student's (Nathan's) initial model of the motion of balls on three ramps, none of the balls were modeled as accelerating; the ball on the steepest slope had the largest speed, while the ball on the ground had the lowest speed. When one of the facilitators noticed this, he asked Nathan to revisit the physical setup, and to observe carefully whether the balls were all starting out with some (non-zero) speed. It was through this iterative comparison between his computer model and the real-world scenario, that Nathan realized that initial conditions play an important role in kinematics. In this case, in order to model the scenario shown in the physical setup, Nathan realized that the yellow and red balls should start from rest. We found that in both the studies, nearly all students at first modeled the blue ball to move at a constant speed; it was only after several rounds of validation of their computational model by re-examining the real-world scenario a few times, that they realized that the blue ball in their simulation should slow down every step. This shows that validation of the simulated world with the real world played an important role in the iterative refinement of students' models.

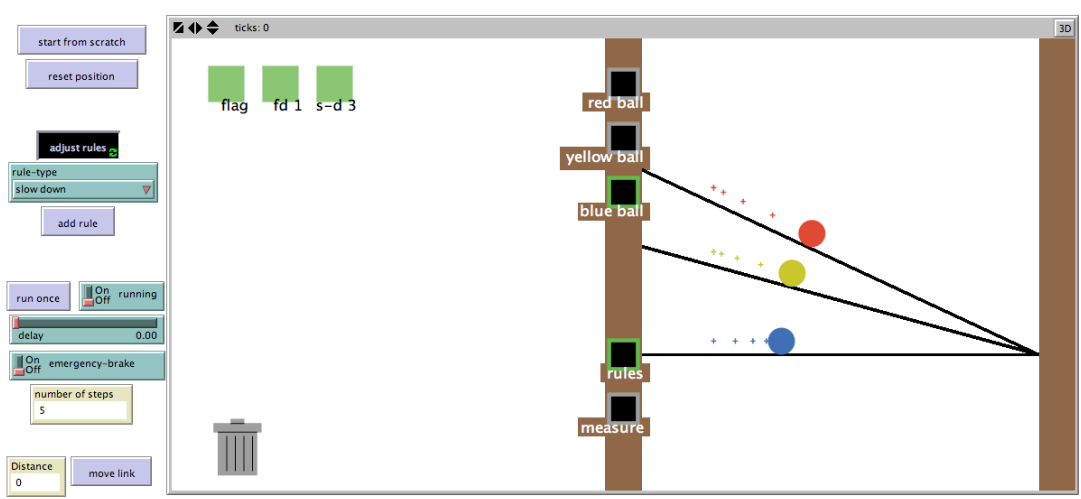

Figure 7. ViMAP-MoMo.

**Phase 3: Graphing Motion**

    **Design Rationale**

From the perspective of the development of students' conceptual understanding of physics, the activities in Phase 3 are intended to help students view the process of changing speeds of the different balls as a continuous phenomenon. Dykstra and Sweet (2009) describe an intuitive "snapshot" view of motion, which affords students a discrete view of motion at any instant. The flagged representation (based on the Logo-based dot-trace representation) that we described in the previous section is designed to leverage this view. In this phase, students learn to piece together multiple snapshots—where each snapshot corresponds to movement of the wabbit(s) during a single time-interval (or iteration)—and develop a view of change in speed and position as a continuous process. From the perspective of development of students' computational thinking and modeling expertise, the activities in this phase involve students learning to analyze the models they developed in Phase 2 using new forms of mathematical measures—speed vs. time and distance vs. time graphs of the motion of each ball.

    **Activities**

    *Activity 3.1: Graphing motion*

This phase introduces a new utility of ViMAP: measure. In this phase, students begin their investigation by inventing measures of motion based on their previously built model from Phase 2. They can access the measure utility simply by clicking on the icon labeled "measure" in the ViMAP interface, which clears the C-World by hiding the command blocks for each ball. The learner can toggle back and forth between the code window and the measure window by clicking on the "rules" and "measure" tab respectively. We describe below how students can use the measure utility to develop graphs of motion.

With the measure tab selected, if the learner then clicks on any two flags in the E-World, then the flags are immediately highlighted and a link (straight line) appears, connecting the two flags (see Figure 8, right side). The learner then clicks the "move link" button to generate a copy of the link (oriented vertically) of the same color as the ball being investigated, and with the same height as the distance between the clicked flags. These vertical bars are generated in the C-World, and they also have labels denoting the height of the bar, which in turn represents the distance between the parent flags.   Figure 8 illustrates a typical screen showing the measure window during this activity. Notice that the left side of the interface is now occupied by series of links (vertical bars) that are labeled by their lengths.   Once the speed and distance bars are created in the C-World, the learner can click, drag, and move the links in order rearrange them and identify trends in the change (over time) of speeds and distances of the balls.

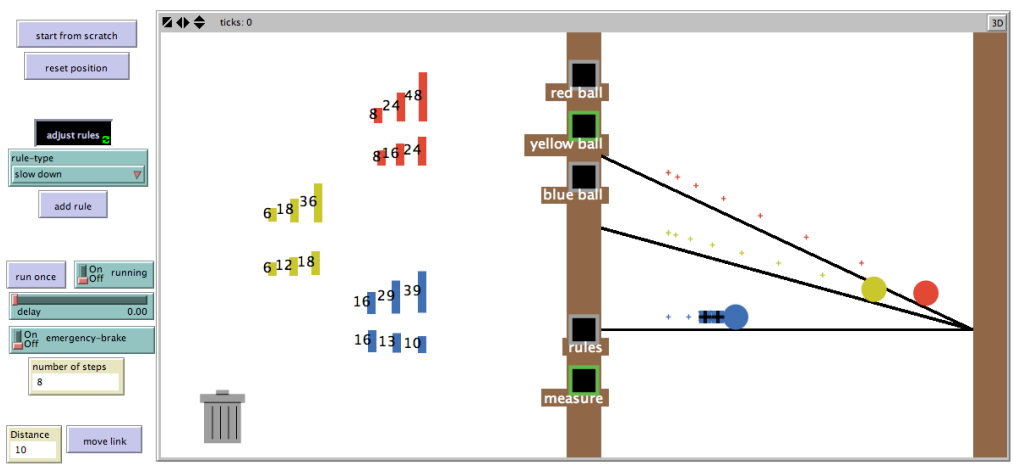

Figure 8. Graphing motion of balls on inclined planes using the "measure" functionality

Students can generate two types of bar graphs: distance vs. time (top graph for each ball in Figure 8), and speed vs. time (bottom graph for each ball in Figure 9). In order to generate a distance vs. time graph, students need to first create distance bars by clicking on an initial flag and any other flag, and then pressing the "move link" button. In order to generate a speed bar that indicates instantaneous speed (i.e., the speed of a wabbit between one instant (or iteration) and another), students need to create distance bars between any two successive flags. It is important to note that these plots afford the opportunity to make explicit to the students the quadratic relationship between distance and time (see top graph in Figure 9), and the linear dependence of speed on time (see bottom graph in Figure 9) in a field of constant acceleration. This is shown in Figure 9.

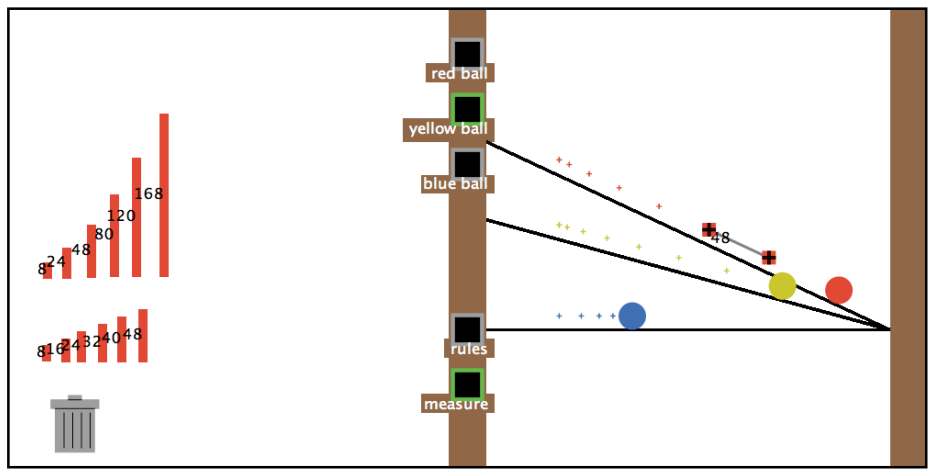

Figure 9. Bar graphs (on the left) make explicit the quadratic dependence of distance on time (top graph) and linear dependence of speed on time (bottom graph)

*Activity 3.2: Reverse engineering speed-time graphs*

In this final activity, students generate shapes and turtle trajectories based on a given speed vs. time graph. For this activity, the instructor will provide all students with a general speed vs.time graph (a sample graph is shown in Figure 10a). Students use the version of ViMAP-MoMo used in Phase 1 including the measure utility, to construct models in which the overall motion of the wabbit reflects a trend in the change of its instantaneous speed (Figure 10b) that is similar to the speed vs. time graph that will be provided to them (Figure 10a). As explained in the earlier activity, each step of the turtle can be selected by the learner and recorded as a speed bar using the "measure" and "move link" functionalities. The nature of this complex task allows for a large number of possible solutions that can leverage students' creativity. After construction of the model, with iterative runs and debugging, students "check" their model by arranging the links sequentially to see if the relationships among instantaneous speeds are similar to the original speed-time graph. Student may choose to display the labels of the links. Figure 10 shows that the general patterns of change in the height of the links that corresponds to the individual displacements in every time-step (as shown in 10c) matches the pattern of the graph (as shown in 10a).

We see these activities as particularly promising for helping students understand that the curvilinear nature of a speed-time graph does not tell us about the "shape" of the motion of the agent (e.g., the car drove down a hill – which has been noted as a common naïve interpretation of the graph shown in Figure 10a), but instead shows us a composite picture of trends in speed over a period of time.

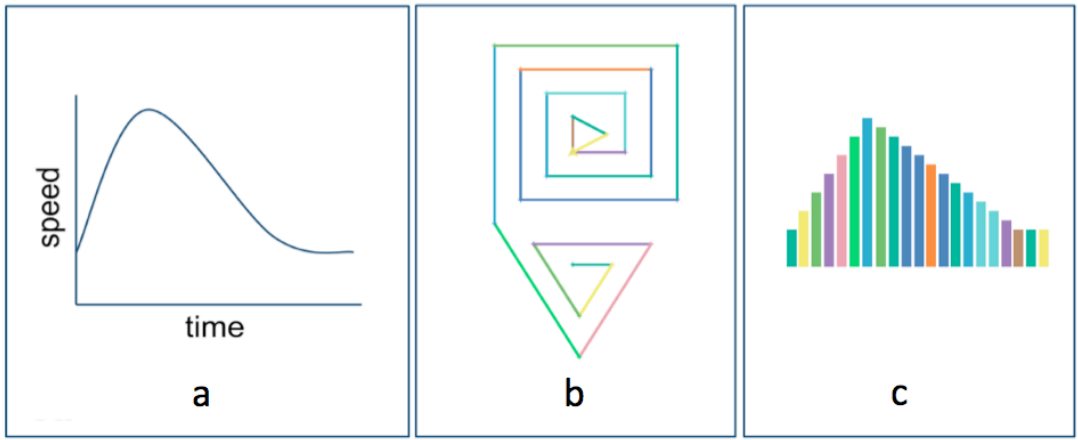

Figure 10. The initial speed-time graph (a), the trace of the wabbit's enacted motion (b), and the bar graph generated from the wabbit's motion (c).

## Summary and Discussion

To summarize, in this paper, we have highlighted the following affordances of ViMAP. First, we showed how ViMAP builds on and extends the principles of concreteness and spatiality that underlay the design of Boxer. In terms of the diSessa's notion of concreteness, i.e., making all the mechanisms of the system visible and directly manipulable by the students, ViMAP enables students to compose and see the program at the same time as the running environment. However, ViMAP goes a step further by making the user's code "live",i.e., the program highlights the command that is being executed, and updates the relevant parameters in the C-World as the system runs. In addition, ViMAP enables students to go beyond programming and generate graphs within the same environment, and they can toggle back and forth between the graph and the ViMAP code. Students are thus introduced to multiple computational and mathematical representations of the simulated phenomenon within the same learning environment. The principle of spatiality is extended by using visual programming, as opposed to text-based programming, as the mode of construction of algorithms, as well as generation of graphs. Another important affordance of ViMAP is the combination of domain-specific and domain-general programming commands. Besides using domain-general commands for control flow, the ViMAP language also contains domain-specific programming commands such as speed-up and slow-down. We believe that this plays an important role in reducing the overhead involved in learning Logo programming in a physics classroom.

Finally, we have described in detail a set of learning activities, sequenced in three distinct but related phases, as well as the rationale underlying these phases. We believe that these activities elucidates a pedagogical approach that integrates Logo-programming and agent-based modeling seamlessly with a kinematics curriculum, and reduces the overhead associated with learning programming prior to learning physics. Another contribution of our paper is that we have also combined the teaching of the physics with aesthetic goals, which we believe is an unique feature of our approach. As we argued earlier, following Miller (1978) and Weschler (1978), our emphasis on aesthetics is reliant on the form of the graphics, particularly the symmetry, continuity, and discontinuity of the Logo graphics in the E-World. The activities and the computational primitives we designed tie the process of shape generation to domain specific learning goals. The aesthetic attention that students provide to the shapes is not distal from their attention to modeling motion. As Weschler (1978) suggested, we see the aesthetic considerations that are taken up in the processes of modeling as potentially deterministic to the "form, development, and efficacy of models" (p. 3), much akin to the development of scientific theories (Miller, 1978).

We conclude this paper with an useful play on a particular word that holds a central place in the constructionist literature: *concreteness*. We highlight three related uses of the term by Papert (1980), diSessa (1985; diSessa, Abelson & Polger, 1991), and Wilensky (1991), as each use highlights a key characteristic of the learning environment we introduced in this paper. Moving beyond Piaget's distinction of concrete operations from formal thinking (Piaget, 1957), Papert (1980) argued that the protean Logo turtle and the Logo programming language provide a way to concretize and personalize the formal. DiSessa further extended the principle of concreteness (diSessa, 1985; diSessa, Abelson & Polger, 1991) to indicate that all the mechanisms of a system must be visible to the learner at all times. As we have argued earlier, diSessa's notion of concreteness is a key rationale behind the design of the ViMAP environment, including the flexibility of the ViMAP programming language itself. Wilensky's notion of concreteness offers us an effective way of articulating the rationale behind the design of ViMAP-MoMo (including the ViMAP environment, programming commands and learning activities) from the perspective of learning kinematics. According to Wilensky, concreteness is the property that measures a person's relationship to an object in terms of the richness of the person's representations of and interactions and connections with the object. The learning activities that we have reported here involve generating multiple forms of computational and mathematical representations of key concepts and phenomena in the domain of kinematics, albeit using a limited vocabulary of programming commands. Many of the activities we presented are non- canonical, but support the development of authentic epistemic practices such as modeling and graphing (from the perspective of learning the physics involved), and reasoning at multiple levels of abstraction (from the perspective of learning to think computationally). As students progress through the sequence of these activities, they begin to develop new meanings and ways of generating inscriptions, and use them to represent kinematic phenomena in non-canonical ways (e.g., using shape drawing to model continuous changes in speed over a period of time). We argued earlier that some of these representations have the potential to be engaging and personally meaningful to the learners. Through participation in these visual programming and agent-based modeling activities, we therefore believe that novice physics learners in a wide age range—elementary to high school—can develop concrete relationships with key aspects in the domain of introductory Newtonian mechanics (both concepts and epistemic practices), and in particular, develop a concrete understanding of motion as continuous change.

## Acknowledgements

The authors gratefully acknowledge the support of Wilson Hubbell and Gokul Krishnan for sharing and working with important ideas during formative stages of this work. The first author would especially like to thank Rich Lehrer for encouraging this work since its ideation. Partial financial support was provided by Vanderbilt University and a grant from the National Science Foundation (NSF IIS # 1124175).